\documentclass[12pt]{iopart}
\usepackage{epsfig}
\def\lesssim{\mathrel{\hbox{\rlap{\hbox{\lower4pt\hbox{$\sim$}}}\hbox{$<$}}}}
\def\gtrsim{\mathrel{\hbox{\rlap{\hbox{\lower4pt\hbox{$\sim$}}}\hbox{$>$}}}}
\def\Box{\opensquare}

\begin{document}

\title[Convergence of Scalar-Tensor theories]{Convergence of Scalar-Tensor theories toward 
General Relativity and Primordial Nucleosynthesis}

\author{A. Serna\dag, J.-M. Alimi\ddag, and A. Navarro\P}

\address{\dag\ Dept.  F\'\i sica y Computaci\'on,  Universidad  Miguel
Hern\'andez,  , E03202-Elche,  Spain}
\address{\ddag\ LAEC (CNRS-UMR  8631), Observatoire de
Paris-Meudon,   F92195-Meudon, France}
\address{\P\ Dept.   F\'\i    sica, Universidad de Murcia, E30071-Murcia, Spain}

\begin{abstract}
In  this paper, we   analyze the   conditions for  convergence  toward
General   Relativity of scalar-tensor  gravity  theories defined by an
arbitrary coupling function $\alpha$ (in the  Einstein frame). We show
that, in  general, the  evolution  of the scalar field  $(\varphi)$ is
governed by two opposite  mechanisms:  an attraction  mechanism  which
tends to  drive scalar-tensor models toward   Einstein's theory, and a
repulsion mechanism  which  has the  contrary effect.  The  attraction
mechanism dominates the recent  epochs  of the universe evolution  if,
and  only if,  the  scalar field   and its derivative  satisfy certain
boundary conditions. Since   these conditions for convergence   toward
general relativity depend on  the particular scalar-tensor theory used
to describe the universe evolution,  the nucleosynthesis bounds on the
present value of  the  coupling function, $\alpha_0$, strongly  differ
from  some  theories to  others. For  example, in  theories defined by
$\alpha  \propto \mid\varphi\mid$ analytical   estimates lead to  very
stringent nucleosynthesis     bounds    on      $\alpha_0$  ($\lesssim
10^{-19}$). By contrast,  in scalar-tensor theories defined by $\alpha
\propto \varphi$ much larger limits on $\alpha_0$ ($\lesssim 10^{-7}$)
are found.
\end{abstract}
\pacs{04.50.+h, 04.80.+z, 98.80.Cq, 98.80.Hw}
\maketitle

\section{Introduction} \label{introduction}

Scalar-tensor (ST) gravity  theories \cite{B,W,N} have  become a focal
point  of interest  in  many   areas  of  gravitational  physics   and
cosmology.   They provide the  most natural generalizations of General
Relativity (GR) by introducing an additional scalar field, $\phi$, the
dynamical  importance of which is determined  by an arbitrary coupling
function $\omega(\phi)$.  Indeed,  most   recent attempts  at  unified
models  of   fundamental    interactions,  i.e.,   string     theories
\cite{Green88} predict the existence of scalar  partners to the tensor
gravity of GR, and would have ST gravity as their low energy limit. In
addition, ST theories are  important in cosmology because they provide
a natural (non-fine tuned) way of exiting the inflationary epoch.

Solar system experiments  (time delay in  the Viking Mars data,  Lunar
Laser Ranging, etc.)  can  put limits on the  present  deviation of ST
theories with   respect to GR.  It is   admitted \cite{Will} that such
experiments impose the constraint $\omega_0>500$  to the present value
of the  coupling function  and,  therefore, that  ST theories   are at
present very close to   General Relativity.  This limit  on $\omega_0$
does  not  necessarily imply  that the universe  evolution  is, at any
time, very   close    to that  found in   GR.   It  has   been   shown
\cite{DN93,SKW99}   that, in   some  ST   theories,  the  cosmological
evolution  drives the scalar   field toward a state  indistinguishable
from GR. Within this 'attraction mechanism', the scalar field can play
an  important  role only in   early cosmology because,  afterwards, it
evolves toward a state with a vanishingly small scalar contribution.

The dynamical importance of the scalar field in the early universe can
be checked by means of the observed abundance of light elements, which
has to  be explained as  an outcome  of the primordial nucleosynthesis
process (PNP).   Contrary to the  weak   field limit, there is  not  a
commonly   accepted  PNP  constraint   on $\omega_0$.     Some authors
\cite{Santiago97,DP99} have recently  found that, in  the framework of
some ST  theories, $\omega_0\gtrsim10^7$  is  required  to obtain  the
observed primordial abundance  of    light elements.  Other    authors
\cite{DTYepes87,SA96b} have instead found  that the PNP test typically
imposes the limit $\omega_0\gtrsim10^{20}$.

In this   paper we will first  reexamine  the attraction  mechanism of
Refs. \cite{DN93,SKW99} and, then,  we will investigate the reason for
the enormous  discrepancy (thirteenth orders of  magnitude) in the PNP
bound on $\omega_0$ obtained by  different authors. We will show that,
in general,  the evolution  of the  scalar  field is  governed  by two
opposite   mechanisms: an attraction  and  a  repulsion mechanism. The
attraction  mechanism  dominates  the recent   epochs  of the universe
evolution only if the scalar field and  its derivative satisfy certain
boundary  conditions  which  depend on   the  particular scalar-tensor
theory   used to describe the  universe  evolution. We will also apply
this   generalized formalism  to   the   theories considered   in Ref.
\cite{SA96b},  where  the coupling  function  was  restricted  to be a
monotonic function of time. We will show that the
nucleosynthesis   bounds  numerically    obtained  in \cite{SA96b}  ($
\omega_0\gtrsim 10^{20} $) are in  close agreement with the analytical
estimates for these  theories. When the same  arguments are applied to
other  ST theories (as  those of Refs. \cite{DN93,SKW99}), one obtains
much less stringent bounds on $\omega_0$. Consequently, the particular
ST  theory  used to  describe   the universe  evolution has  a crucial
importance on the PNP  limit on $\omega_0$  and, therefore, it is  not
possible to establish a general and unique limit for all ST models.

The  plan   of  the paper  is  as   follows. We   begin  outlining the
scalar-tensor theories as well as the two frames usually considered to
build     up       cosmological   models     in     their    framework
(Sec. \ref{section-ST}). An autonomous    evolution equation for   the
scalar field is  then obtained in   Sec. \ref{Decoupled} both  for the
Jordan and the Einstein frame. Using this equation, and the particular
family of scalar-tensor theories specified in Sec. \ref{The-coupling},
we  then  analyze  the  evolution of   the scalar  field both  in  the
radiation-dominated  epoch  (Sect.  \ref{Radiation-epoch})   and   the
matter-dominated  epoch  (Sect. \ref{Matter-epoch}).  Our  results are
then applied to estimate  the nucleosynthesis bounds on $\omega_0$ for
this  family   of theories    (Sec. \ref{nucleosynthesis}).   Finally,
conclusions   and     a   summary  of  our      results are   given in
Sec. \ref{conclusions}.

\section{Scalar-Tensor Gravity Theories} \label{section-ST}

\subsection{The Jordan frame}

We consider a scalar-tensor  gravity theory in which the gravitational
interaction is carried by the  metric  $g_{\mu\nu}$ and an  additional
massless scalar field, $\phi$. Since these fields are measured through
laboratory  clocks   and  rods,  which  are  made    of atoms and  are
essentially   based   on  a   non-gravitational    physics, the action
describing a scalar-tensor   gravity  theory must keep unaltered   the
basic  laws  of non-gravitational interactions  (as, e.g., statistical
physics and  electrodynamics). When this  fact  is taken into account,
the resulting  space-time  units are termed   as "atomic", "Jordan" or
"physical" units \cite{Dirac73,Canuto77,HolmanKW90,HolmanKVW91}.

Using the Jordan frame, the  most general action describing a massless
scalar-tensor theory of gravitation is \cite{B,W,N}

\begin{equation}
S = \frac{1}{16\pi}\int (\phi{\cal R} -\frac{\omega(\phi)}{\phi}
\phi_{,\mu} \phi^{,\mu}) \sqrt{-g} d^{4}x + S_{M} \label{action}
\end{equation}

\noindent
where ${\cal R}$ is the curvature scalar of the metric $g_{\mu\nu}$,
$g \equiv   \mbox{det}(g_{\mu\nu})$, $\phi$ is the scalar   field, and
$\omega(\phi)$ is an arbitrary coupling function.

The variation of Eq.\ (\ref{action})  with respect to $g_{\mu\nu}$ and
$\phi$ leads to the field equations:
\numparts
\begin{eqnarray}
 {\cal R}_{\mu\nu}-\frac{1}{2}g_{\mu\nu} {\cal R} &=& -
 \frac{8\pi}{\phi} T_{\mu\nu} - \frac{\omega}{\phi^{2}}(\phi_{,\mu}
 \phi_{,\nu} - \frac{1}{2} g_{\mu\nu} \phi_{,\alpha}
 \phi^{,\alpha})\nonumber\\ & & -\frac{1}{\phi} (\phi_{,\mu;\nu} -
 g_{\mu\nu} \Box \phi )
\label{eq:field1} \\
(3+2\omega)\Box\phi &=& 8\pi T - \frac{d\omega}{d\phi}
\phi_{,\alpha}\phi^{,\alpha} \label{eq:field2}
\end{eqnarray}
\endnumparts
which satisfy the usual conservation law

\begin{equation}
T^{\mu\nu}_{;\nu} = 0
\end{equation}

\noindent
where $T^{\mu\nu}$ is the energy-momentum tensor and $\Box\phi\equiv
g^{\mu\nu}\phi_{,\mu;\nu}$.

For  a  homogeneous  and  isotropic universe,  the  line-element has a
Robertson-Walker form   and the energy-momentum  tensor corresponds to
that of   a perfect fluid.  The field  equations (\ref{eq:field1}) and
(\ref{eq:field2}) then become
\numparts
\begin{eqnarray}
&&\frac{8\pi}{3\Phi}\rho = \frac{c^{2}K}{R^{2}}+
\frac{\dot{R}^{2}}{R^{2}} -
\frac{\omega}{6}\frac{\dot{\Phi}^{2}}{\Phi^{2}}+
\frac{\dot{R}\dot{\Phi}}{R\Phi}\\
&&-\frac{8\pi G}{3\Phi}(\rho+3P/c^2) =
2\frac{\ddot{R}}{R}+\frac{\dot{R}}{R}\frac{\dot{\Phi}}{\Phi}
+\frac{\ddot{\Phi}}{\Phi}+\frac{2\omega}{3}\frac{\dot{\Phi}^2}
{\Phi^2}\\
&&\ddot{\Phi}+3\frac{\dot{R}}{R}\dot{\Phi} =
\frac{1}{(3+2\omega)}[8\pi G(\rho-3P/c^{2})-
\frac{d\omega}{d\Phi}\dot{\Phi}^2]
\end{eqnarray}
\endnumparts

\noindent where $K=0, \pm 1$, $ \Phi \equiv G\phi $, $R(t)$ is the scale
factor,  $\rho$ and  $P$  are  the energy-mass  density and  pressure,
respectively, and dots mean time derivatives. In addition, we have the
usual conservation equation:

\begin{equation}
d(\rho R^{3}) +(P/c^{2}) dR^{3} = 0 \label{eq:conserv}
\end{equation}

\noindent which ensures that the standard laws of non-gravitational
interactions are not modified by the presence of a scalar field.

\subsection{The Einstein frame}

When the metric is assumed to be measured through purely gravitational
clocks  and rods,  the space-time  units  are termed as  "Einstein" or
"spin" units. In this frame, the general  action describing a massless
scalar-tensor  theory can be  obtained   from Eq. (\ref{action}) by  a
conformal transformation
\numparts
\begin{eqnarray}
g_{\mu\nu} = A^{2}(\varphi) g_{\mu\nu}^{\ast} \label{eq:A1}\\
A^{2}(\varphi) = (\Phi)^{-1} \label{eq:A2}
\end{eqnarray}
\endnumparts

\noindent where $A(\varphi)$ is an arbitrary function related to
$\omega(\Phi)$ by

\begin{equation}
 \alpha = (3+2\omega)^{-1/2}=\frac{d\ln A}{d \varphi}
\label{eq:alpha}
\end{equation}

Using Eqs. (\ref{action}), (\ref{eq:A1}), (\ref{eq:A2}) and (\ref{eq:alpha}), one
obtains

\begin{equation}
S_{\ast} = \frac{c^4}{16\pi G_{\ast}}\int ({\cal R}_{\ast} -
2\varphi_{,\mu} \varphi^{,\mu}) \sqrt{-g_{\ast}} \frac{d^{4}x}{c} +
S_{M}^{\ast} \label{eq:SE}
\end{equation}

\noindent where $G_{\ast}$ is
Newton's constant and asterisks denote
quantities expressed in Einstein units.   Since our measures are based
on non-purely gravitational rods and clocks, quantities written in the
Einstein   frame are not  observable.   Comparison  between theory and
observations must be then performed by using the Jordan frame.

From the action (\ref{eq:SE}), the Einstein field equations are:
\numparts
\begin{eqnarray}
 {\cal R}_{\mu\nu}^{\ast} &=& 2\varphi_{,\mu}\varphi_{,\nu}+
8\pi G_{\ast}(T_{\mu\nu}^{\ast} - \frac{1}{2}T^{\ast}g_{\mu\nu}^{\ast})
\label{eq:Efield1} \\
\Box_{\ast}\varphi &=& -4\pi G_{\ast}\alpha(\varphi)T_{\ast}
\label{eq:Efield2}
\end{eqnarray}
\endnumparts

\noindent where $ T_{\mu\nu}^{\ast}=2g_{\ast}^{-1/2}\delta S_m/\delta
g_{\mu\nu}^{\ast}$ is the energy-momentum tensor in Einstein
units.

If   we consider  a  homogeneous   and  isotropic universe,  the field
equations (\ref{eq:Efield1})-(\ref{eq:Efield2}) become
\numparts
\begin{eqnarray}
& & -3\frac{1}{R_{\ast}}\frac{d^2 R_{\ast}}{dt^{2}_{\ast}} =
4\pi G_{\ast}(\rho_{\ast}+3P_{\ast}) +
2 \left(\frac{d\varphi}{dt_{\ast}}\right)^2
\label{eq:Eeq1} \\
& & 3\frac{1}{R^{2}_{\ast}}\left(\frac{dR_{\ast}}{dt_{\ast}}\right)^2
+3\frac{K}{R_{\ast}^2} = 8\pi G_{\ast} \rho_{\ast}+
\left(\frac{d\varphi}{dt_{\ast}}\right)^2
\label{eq:Eeq2}\\
& & \frac{d^2\varphi}{dt^{2}_{\ast}}+
3\frac{1}{R_{\ast}}\frac{dR_{\ast}}{dt_{\ast}}\frac{d\varphi}{dt_{\ast}}
= -4\pi G_{\ast} \alpha(\varphi)(\rho_{\ast}-3P_{\ast})
\end{eqnarray}
\endnumparts
and the 'conservation' equation is modified to

\begin{equation}
d(\rho_{\ast}R_{\ast}^3)+P_{\ast}d(R_{\ast}^3)=(\rho_{\ast}-3P_{\ast})
R_{\ast}^3da(\varphi)
\end{equation}

Since the mass-energy is not conserved in Einstein units, the basic
laws of non-gravitational physics are modified in this  frame (see
\cite{DTS90,SDT92,Serna96} for a  reformulation of nuclear reaction
rates and thermodynamics in Einstein units).

\section{Decoupled evolution of the Jordan scalar field}
\label{Decoupled}
In  the form given  above, the time evolution  of the scale factor and
the scalar  field are  coupled both in   the Jordan and   the Einstein
frame.  Previous    works  \cite{DN93,SKW99}  have   found   that,  by
introducing an appropriate change of variables, it is possible to find
an    evolution equation  for the    Einstein   scalar field  which is
independent of the cosmic scale factor. We will now show    that it
is also  possible   to find  such a decoupled
evolution equation for the Jordan  scalar field.

Let us define the functions
\numparts
\begin{eqnarray}
&&\psi\equiv \frac{1}{2}\ln\Phi\\
&&\gamma \equiv \frac{P/c^2}{\rho}\\
&&\epsilon \equiv \frac{3c^2 K\Phi}{8\pi G\rho R^2}\\
&&W \equiv (3+2\omega)/3
\end{eqnarray}
\endnumparts
The Jordan evolution equations of $R$ and $\Phi$ then become:
\numparts
\begin{eqnarray}
&&\frac{\ddot{R}}{R}+\frac{\dot{R}}{R}\dot{\psi}+\ddot{\psi}+2W\dot{\psi}^2
=-\frac{4\pi G\rho}{3\Phi}(1+3\gamma)\label{Jeqs1}\\
&&\frac{8\pi G\rho}{3\Phi}(1-\epsilon)
=\left(\frac{\dot{R}}{R}+\dot{\psi}\right)^2-W\dot{\psi}^2\label{Jeqs2}\\
&&2\ddot{\psi}+4\dot{\psi}^2+6\frac{\dot{R}}{R}\dot{\psi}
=\frac{8\pi G\rho}{3\Phi
W}(1-3\gamma)-\frac{1}{W}\frac{dW}{d\psi}\dot{\psi}^2\label{Jeqs3}
\end{eqnarray}
\endnumparts

In order to obtain a decoupled evolution  equation for $\psi$, we will
define a 'time' parameter, $p$, as:
\begin{equation}\label{pdef}
dp=h_c dt; \qquad h_c=\frac{\dot{R}}{R}+\dot{\psi}
\end{equation}
In  terms   of   these  variables, and    denoting   $f'\equiv df/dp$,
Eqs. (\ref{Jeqs1})-(\ref{Jeqs3}) reduce to:
\begin{eqnarray}
&&\frac{h^{\prime}_{c}}{h_c}=
\psi'-1-2W\psi'^2-\frac{4\pi G\rho}{3\Phi h^{2}_{c}}(1+3\gamma)
\label{hcp}\\
&& h^{2}_{c}(1-W\psi'^2)=
\frac{8\pi G\rho}{3\Phi}(1-\epsilon)\label{hc2}\\
&&\psi''+\frac{h^{\prime}_{c}}{h_c}\psi'-\psi'^2+3\psi'=\nonumber\\
&&\qquad\quad=
\frac{4\pi G\rho}{3\Phi h^{2}_{c}}\frac{1-3\gamma}{W}-
\frac{1}{2W}\frac{dW}{d\psi}\psi'^2\label{psipp}
\end{eqnarray}
and,  using Eqs. (\ref{hc2})  and  (\ref{hcp}) to eliminate $h_c$  and
$h^{\prime}_{c}$, respectively, we finally obtain:
\begin{eqnarray}
\frac{2(1-\epsilon)}{1-W\psi'^2}\psi''&+&(3-3\gamma-4\epsilon)\psi'
=\nonumber\\
&=& \frac{1-3\gamma}{W}-\left(\frac{1-\epsilon}{1-W\psi'^2}\right)
\frac{1}{W}\frac{dW}{d\psi}\psi'^2\label{Jdecoupled}
\end{eqnarray}
which is an evolution  equation for $\psi=(1/2)\ln\Phi$ independent of
the evolution  of the   cosmic scale factor.   This  equation  is  the
analogous,      in  the     Jordan   frame,   to     Eq.  (3.15)    of
\cite{DN93}. Therefore, we find   that  a decoupled evolution   of the
scalar  field is not  an exclusive  feature of  the Einstein frame. 

In order to  find the well-known decoupled evolution  equation for  the Einstein scalar
field,     we   will   first  multiply    Eq.    (\ref{Jdecoupled}) by
$W^{1/2}(1-W\psi'^2)/2(1-\epsilon)$. We obtain:
\begin{eqnarray}
\frac{2(1-\epsilon)}{1-W\psi'^2}&&(W^{1/2}\psi')'=\nonumber\\
&&=\frac{1-3\gamma}{W^{1/2}}-
3 (1-\gamma-\frac{4}{3}\epsilon)W^{1/2}\psi'
\label{Jdecoupled2}
\end{eqnarray}

We can now  change to the Einstein  frame by performing  the conformal
transformation (\ref{eq:A1})-(\ref{eq:A2}) with
\begin{equation}\label{varphipsi}
\frac{d\varphi}{d\psi}=-\sqrt{3W}=-\alpha^{-1}
\end{equation}

Eq. (\ref{Jdecoupled2}) then becomes:
\begin{equation}
\frac{2(1-\epsilon)}{3-\varphi'^2}\varphi''+
(1-\gamma-\frac{4}{3}\epsilon)\varphi'=-\alpha (1-3\gamma)
\label{Edecoupled}
\end{equation}
which agrees with    Eq.  (3.15)   of   \cite{DN93}.

We also note that, by defining $H_{\ast}$ and $\rho_{\ast}$ through
\begin{equation}\label{Hast}
\Phi^{1/2}H_{\ast}=h_c;\qquad \rho=\Phi^2\rho_{\ast},
\end{equation}
Eq. (\ref{hc2}) yields
\begin{equation}\label{Hast2}
\frac{8\pi G\rho_{\ast}}{H^{2}_{\ast}}=
\frac{3-\varphi'^2}{1-\epsilon}
\end{equation}

From this equation we see that, when $\epsilon=0$, the
local positivity of the energy density implies that
\begin{equation}
\varphi'^2\le 3.
\end{equation}

Most papers  analyzing  the convergence  toward General Relativity  of
scalar-tensor theories  are based on the  Einstein frame.  In order to
make   easier  the comparison  of  our  results  with  those founds in
previous    works,     we  will     use   hereafter    the    Einstein
frame. Nevertheless, it must be noted that most of our conclusions can
also   be   found   by    using   the  Jordan    evolution    equation
(\ref{Jdecoupled}).

\section{The coupling function}\label{The-coupling}

The time evolution of scalar-tensor theories can be studied only after
specifying  a functional form  of   the coupling function. Barrow  and
Parsons \cite{BP97}    have noted that most    expressions used in the
literature for  ($3+2\omega$) can  be classified into  three different
families of theories:
\numparts
\begin{eqnarray}
&&\mbox{a) Theories-1:}\quad 3+2\omega=\frac{1}{B_1|\Phi-1|^\delta}
\quad(\delta>1/2)\label{theories1}\\
&&\mbox{b) Theories-2:}\quad 3+2\omega=\frac{1}{B_1|\ln\Phi|^\delta}
\;\quad(\delta>1/2)\label{theories2}\\
&&\mbox{c) Theories-3:}\quad 3+2\omega=\frac{1}{B_1|\Phi^\delta-1|}
\quad(\delta>0)\label{theories3}
\end{eqnarray}
\endnumparts

\noindent where $B_1$ is an arbitrary positive constant.

The three  theories    defined  by Eqs. (\ref{theories1})-(\ref{theories3})   imply   very
different behaviours  of the early universe,  which have been analyzed
in detail  by   Barrow and Parsons  \cite{BP97}.   The first class  of
theories  has also been  studied  by  Garc\'\i a-Bellido and  Quir\'os
\cite{GBQ90},    Serna    and    Alimi    \cite{SA96a},    Comer    et
al. \cite{Comer97}, and Navarro   et al. \cite{NSA99}. However, it  is
important  to note that all these  theories have similar behaviours in
the limit close   to General  Relativity  ($\Phi\rightarrow1$).  As  a
matter of fact,  since  $\ln\Phi\simeq \Phi -  1$  and $\Phi^\delta -1
\simeq\delta\ln\Phi$, we can take
\begin{equation}
3+2\omega=\frac{1}{B_1|\ln\Phi|^\delta}
=\frac{1}{B_1|2\psi|^\delta} \label{Jcoupling}
\quad(\delta>1/2)
\end{equation}
to represent  the way in which these  three types of theories approach
the limit of GR.

Using   the  coupling    function  given  by   Eq.  (\ref{Jcoupling}),
integration of Eq. (\ref{varphipsi}) yields
\begin{equation}
\varphi=-\frac{\mbox{sign}(\psi)}{(2-\delta)B_{1}^{1/2}}
|2\psi|^{(2-\delta)/2}\quad(\delta<2)\label{varphi}
\end{equation}
where we have normalized the  integration constant so that $\varphi=0$
corresponds to $\Phi=1$    (or $\psi=0$).  Note that,  according    to
Eq. (\ref{varphi}), sign($\varphi)=-$sign($\psi$).

By   introducing  Eq. (\ref{varphi})  into   Eq. (\ref{Jcoupling}), we
obtain the Einstein form of the coupling function:
\begin{equation}
\alpha=B_2|\varphi|^{\frac{\delta}{2-\delta}}=\kappa(\varphi) |\varphi|
\label{Ecoupling}
\end{equation}
where $B_2\equiv B_{1}^{1/(2-\delta)}(2-\delta)^{\delta/(2-\delta)}  >
0$, and
\begin{equation}
\kappa(\varphi)=B_2|\varphi|^\frac{2(\delta-1)}{2-\delta}\label{elastic}
\end{equation}

\section{Radiation-dominated evolution}\label{Radiation-epoch}

In  the radiation-dominated epoch, the  state equation is $ P/c^2=\rho
/3$ (i.e., $  \gamma =1/3$) and  the curvature  effects are negligible
($\epsilon =0$). Consequently, the  evolution equation of the Einstein
scalar field (Eq. \ref{Edecoupled}) is well approximated by:
\begin{equation}\label{RDevol}
\frac{2}{3-\varphi'^{2}}\varphi''+\frac{2}{3} \varphi'=0
\end{equation}
which does not depend on
the functional form of $\alpha(\varphi)$.

The integration of Eq. (\ref{RDevol}) gives:
\begin{equation}\label{RDvelocity}
\varphi'^{2}=\frac{3k^{2}}{e^{2p}+ k^2}
\end{equation}
where  $k$    is   related    to  the     initial  ($p=0$)    velocity
$\varphi^{\prime}_{R}$ through:
\begin{equation}\label{k2}
k^2=\frac{(\varphi^{\prime}_{R})^2 }{3-(\varphi^{\prime}_{R})^2 }
\end{equation}

In terms of $k$, the solution of Eq. (\ref{RDevol}) is:
\begin{equation}\label{RDvarphi}
\varphi =\varphi_{R}-\sqrt{3}\; \mbox{sign}(k)\ln
\left[\frac{\sqrt{1+k^{2}e^{-2p}}+ke^{-p}}{\sqrt{1+k^2}+k} \right]
\end{equation}

\section{Matter-dominated evolution: The attraction-repulsion mechanism}
\label{Matter-epoch}

Let us now   analyze the evolution    of the scalar field  during  the
matter-dominated  era      ($\gamma=0$)    of    a   flat     universe
($\epsilon=0$). The  evolution equation for  the Einstein scalar field
is, in this case:
\begin{equation}
\frac{2}{3-\varphi'^2}\varphi''+\varphi'+\alpha=0\label{master}
\end{equation}

As in  previous works   \cite{DN93,Santiago97}, we will  first  assume
that,  at   some   time (for  instance,   at  the    beginning  of the
matter-dominated  era), the scalar-tensor theory  is not very far from
GR  so   that we can neglect    $\varphi'^2$  against 3  and, in
addition,  the coupling  function  $\alpha(\varphi)$ is represented by
Eq. (\ref{Ecoupling}).

In this case, the evolution equation (\ref{master}) reduces to:
\begin{equation}
\frac{2}{3}\varphi''+\varphi'+\sigma_\varphi \kappa(\varphi)
\varphi=0\label{master2}
\end{equation}
where
\begin{equation}
\sigma_\varphi=\mbox{ sign}(\varphi)
\end{equation}

When $\sigma_\varphi=+1$, the above expression corresponds to
the evolution equation of a damped harmonic oscillator with a variable
elastic coefficient $\kappa(\varphi)$. The first term ($2\varphi''/3$)
represents the   total  force  on  a   fictitious   particle of   mass
$m=2/3$. The second term ($\varphi'$) corresponds  to a friction force
proportional  to the velocity  and, finally, the third term represents
an elastic force with  a  variable coefficient $\kappa(\varphi)$.  The
existence, in  these conditions, of a  damped oscillatory behaviour of
the Einstein  scalar field was first  reported by  Damour and Nordvedt
\cite{DN93} and  it is  usually  termed as  the 'attraction mechanism'
toward General Relativity .

We note however  that,   if   $\sigma_\varphi=-1$, the effective     elastic
coefficient  $\sigma_\varphi \kappa(\varphi)$ is negative and, consequently,
there  exists  a 'repulsion'  mechanism instead an  attraction one. We
will now analyze the matter-dominated evolution of the scalar field by
considering   separately the   cases   $\delta=1$ and  $\delta\ne1$ in
Eq. (\ref{elastic}).

It must be noted that the  class of scalar-tensor theories analyzed in
\cite{DN93}  correspond  to  a   positive  constant  $\sigma_\varphi$   and,
therefore, they are always attractive.

\subsection{Case of a constant elastic coefficient ($\delta=1$)}

When    $\delta=1$,   the      elastic     coefficient   defined    by
Eq. (\ref{elastic})  becomes a   positive constant $B_2$.   The scalar
field evolution equation then reduces to:
\begin{equation}
\varphi''+\frac{3}{2}\varphi'+\frac{3}{2}\sigma_\varphi B_2\varphi=0
\label{case1}
\end{equation}
Since the roots of the characteristic equation are:
\begin{equation}
r_{\pm}=-\frac{3}{4}\pm\frac{3}{4}\sqrt{1-\frac{8}{3}B_2\sigma_\varphi}
\end{equation}
the  general solution  of  Eq.   (\ref{case1}) admits four   different
behaviours, depending on $B_2$ and $\sigma_\varphi$:

a) Damped harmonic motion ($\sigma_\varphi>0$, $B_2>3/8$)
\begin{equation}
 \varphi = C_1 e^{-\frac{3}{4}p}\cos(\omega_1 p+C_2) \label{solution1}
\end{equation}
where
\numparts
\begin{eqnarray}
&&\omega_1=\frac{3}{4}\sqrt{\frac{8}{3}B_2-1}\\
&&C_1=\varphi_0\left[\left(\frac{\varphi^{\prime}_0+\frac{3}{4}\varphi_0}
{\omega_1\varphi_0}\right)^2+1\right]^{1/2}\\
&&C_2=-\mbox{atan}\left(
\frac{\varphi^{\prime}_0+\frac{3}{4}\varphi_0}{\omega_1\varphi_0}\right)
\end{eqnarray}
\endnumparts

b) Critically damped motion ($\sigma_\varphi>0$, $B_2=3/8$):
\begin{equation}
\varphi = e^{-\frac{3}{4}p}(C_1 p+C_2) \label{solution2}
\end{equation}
where
\begin{equation}
C_1=\varphi^{\prime}_0+\frac{3}{4}\varphi_0; \quad C_2=\varphi_0
\end{equation}

c) Overdamped attraction motion ($\sigma_\varphi>0$, $B_2<3/8$):
\begin{equation}
\varphi = e^{-\frac{3}{4}p}[C_1 e^{\beta_1 p}+C_2 e^{-\beta_1 p}]
\label{solution3}
\end{equation}
where

\numparts
\begin{eqnarray}
&&\beta_1=\frac{3}{4}\sqrt{1-\frac{8}{3}B_2}<\frac{3}{4}\\
&&C_1=\frac{\varphi^{\prime}_0+(\beta_1+3/4)\varphi_0}
{2\beta_1};\\
&&C_2=-\frac{\varphi^{\prime}_0+(-\beta_1+3/4)\varphi_0}
{2\beta_1}
\end{eqnarray}
\endnumparts

d) Overdamped repulsion motion ($\sigma_\varphi<0$, and any $B_2$):
\begin{equation}
 \varphi = e^{-\frac{3}{4}p}[C_1 e^{\beta_2 p}+C_2 e^{-\beta_2 p}]
\label{solution4}
\end{equation}
where
\numparts
\begin{eqnarray}
&&\beta_2=\frac{3}{4}\sqrt{1+\frac{8}{3}B_2}>\frac{3}{4}\\
&&C_1=\frac{\varphi^{\prime}_0+(3/4+\beta_2)\varphi_0}
{2\beta_2}; \label{constant4a}\\
&&C_2=-\frac{\varphi^{\prime}_0+(3/4-\beta_2)\varphi_0}{2\beta_2}\label{constant4b}
\end{eqnarray}
\endnumparts

Apparently, only  the    first  three solutions   converge toward   GR
($\varphi=0$)   while   the   last one   diverges   to  $\pm\infty$ as
$p\rightarrow \infty$.   However, it must be noted   that each of such
solutions   governs the whole  matter-dominated  era if,  and only if,
$\sigma_\varphi$ does not change. This is the case of the models analyzed in
Refs.  \cite{DN93,SKW99} where, in  Eq.  (\ref{case1}),  $\sigma_\varphi$ is
forced to be $\sigma_\varphi\equiv+1$ at any time.

In the case of the models considered here, where the sign $\sigma_\varphi$ of
$\varphi$ can be variable in time,  the matter-dominated evolution of
the scalar field has a much more complicated behaviour. As a matter of
fact, the above four solutions (Eqs. \ref{solution1}--\ref{solution4})
admit the possibility of a change in the sign of $\varphi$ at a finite
time $p>0$  (this is specially obvious for  the first solution  due to
the oscillatory behaviour   of  the cosinus function).  Therefore,  an
initially convergent model ($\sigma_\varphi=+1$)  could finally diverge from
GR     as    a  consequence   of    a   change     in   the   sign  of
$\varphi$. Reciprocally,  an initially divergent model ($\sigma_\varphi=-1$)
could finally converge toward  GR if sign($\varphi$) changes.  We must
then perform  a more  detailed  analysis  to find  the conditions  for
convergence (or divergence) with respect to GR.

According  to   Eqs.  (\ref{solution1})--(\ref{constant4b}),  $\varphi$
vanishes (and, therefore, $ \sigma _\varphi$ changes)  at a finite  $p>0$
if,   and only  if, the  initial   values of  $\varphi$ and $\varphi'$
satisfy the following conditions:
\begin{equation}
\left\{
\begin{array}{lll}
\mbox{a)} & \mbox{always} & \\
\mbox{b)} & \varphi^{\prime}_{0}<-\frac{3}{4}\varphi_0, & (\varphi_0>0) \\
\mbox{c)} &
\varphi^{\prime}_{0}<-(\frac{3}{4}+\beta_1) \varphi_0, & (\varphi_0>0)\\
\mbox{d)} & \varphi^{\prime}_{0}>-(\frac{3}{4}+\beta_2) \varphi_0,
\quad & (\varphi_0<0)
\end{array}\right.\label{conditions}
\end{equation}

When these conditions are satisfied, an initially attraction behaviour
becomes   a  repulsion  one  at $p_1>0$.   Reciprocally,  an initially
repulsion   behaviour  becomes, at  $p_1>0$,   an  attraction one. The
question is now to determine  whether these new behaviours, reached at
$p>p_1$, govern the rest of the matter-dominated evolution.

Let  us  first  consider the   attraction-to-repulsion  case. In  this
situation,  the attraction   models given by   Eqs. (\ref{solution1}),
(\ref{solution2}) and  (\ref{solution3})  (with $p\le  p_1$)  imply  a
vanishing  scalar   field at  $p=p_1$.  After  this time, the  sign of
$\varphi$ has changed and,  therefore, all models become expressed  by
Eq.  (\ref{solution4})  but with     the integration constants    (see
Eqs. \ref{constant4a} and \ref{constant4b}) given by:
\begin{equation}
C_1=-C_2=\frac{\varphi^{\prime}_{1}}{2\beta_2}\label{a-to-r}
\end{equation}
where $\varphi^{\prime}_{1}$  denotes the $\varphi'$ value at $p=p_1$,
and we have taken into account that $\varphi(p_1)=0$.

Using Eqs.  (\ref{solution4}) and   (\ref{a-to-r}), the  scalar  field
evolution for $p\ge p_1$ is given by:
\begin{equation}
\varphi=\frac{\varphi^{\prime}_{1}}{2\beta_2} e^{-\frac{3}{4}\tilde{p}}
[e^{\beta_2 \tilde{p}}-e^{-\beta_2\tilde{p}}]
\label{diverges}
\end{equation}
where $ \tilde{p}\equiv p-p_1$.

This solution never vanishes  for $p>p_1$ and diverges  to $\pm\infty$
when  $p\rightarrow +\infty$.  Consequently,   none of    these models
finally converges towards GR.

Let    us now consider     the repulsion-to-attraction case.  In  this
situation, the  repulsion model given   by Eq. (\ref{solution4}) (with
$p\le p_1$) implies a  vanishing scalar field  at $p=p_1$. After  this
time, the  sign of $\varphi$ changes  and, therefore, the scalar field
evolution    becomes     expressed     by   Eqs.    (\ref{solution1}),
(\ref{solution2})  or    (\ref{solution3}), depending  on   the  $B_2$
value. The integration constants in this new regime are:
\begin{equation}
\left\{
\begin{array}{lll}
C_1=\frac{|\varphi^{\prime}_1|}{\omega_1}, & C_2=\pi/2 &
(B_2>3/8)\\
C_1=\varphi^{\prime}_1, & C_2=0 &
(B_2=3/8)\\
C_1=\frac{\varphi^{\prime}_1}{2\beta_1}, & C_2=-C_1 &
(B_2<3/8)\\
\end{array}\right.\label{constants3}
\end{equation}
Therefore, when $p\ge p_1$:
\begin{equation}
\varphi=\left\{
\begin{array}{ll}
\frac{|\varphi^{\prime}_1|}{\omega_1}e^{-\frac{3}{4}\tilde{p}}
\sin(\omega_1\tilde{p}) &(B_2>3/8)\\
\varphi^{\prime}_1 e^{-\frac{3}{4}\tilde{p}}\tilde{p} & (B_2=3/8)\\
\frac{\varphi^{\prime}_1}{2\beta_1} e^{-\frac{3}{4}\tilde{p}}
[e^{\beta_1\tilde{p}}-e^{-\beta_1\tilde{p}}] &(B_2<3/8)\\
\end{array}\right.\label{newattract}
\end{equation}

We  then  find that, if   $B_2>3/8$, the new  attraction  behaviour is
oscillatory. Consequently, it will return   later to have a  repulsion
motion similar to that of Eq. (\ref{diverges}),  and these models will
finally    diverge  from GR.    On   the  contrary,  if $B_2\le  3/8$,
Eqs. (\ref{newattract}) imply that $\varphi$ never vanishes at $p>p_1$
except for  $p\rightarrow+\infty$   and, therefore, these  models  are
finally attracted toward GR.

Summarizing,  all models with  $B_2>3/8$ diverge from GR, whatever the
initial  values  of $\sigma_\varphi$, $\varphi$   and $\varphi'$ are. In the
same   way, models with $B_2\le3/8$ and   $\sigma_\varphi>0$ (at $p=0$) will
diverge  from    GR       if conditions    (\ref{conditions})b      or
(\ref{conditions})c are satisfied,  while models with $\sigma_\varphi<0$ (at
$p=0$) will diverge from GR if  conditions (\ref{conditions})d are not
satisfied.  Consequently  (see Fig.   1), the   scalar-tensor theories
defined by Eq. (\ref{Ecoupling})  with $\delta=1$ will converge toward
GR if, and only if $ B_2\leq 3/8$ and:

\begin{equation}\label{Ccondition}
\begin{array}{ll}
 \varphi^{\prime}_{0}\ge   -(3/4+\beta_1)
\varphi_0 & (\mbox{for any $\sigma_\varphi$ value})
\end{array}
\end{equation}

\begin{figure}
\centerline{\epsfig{figure=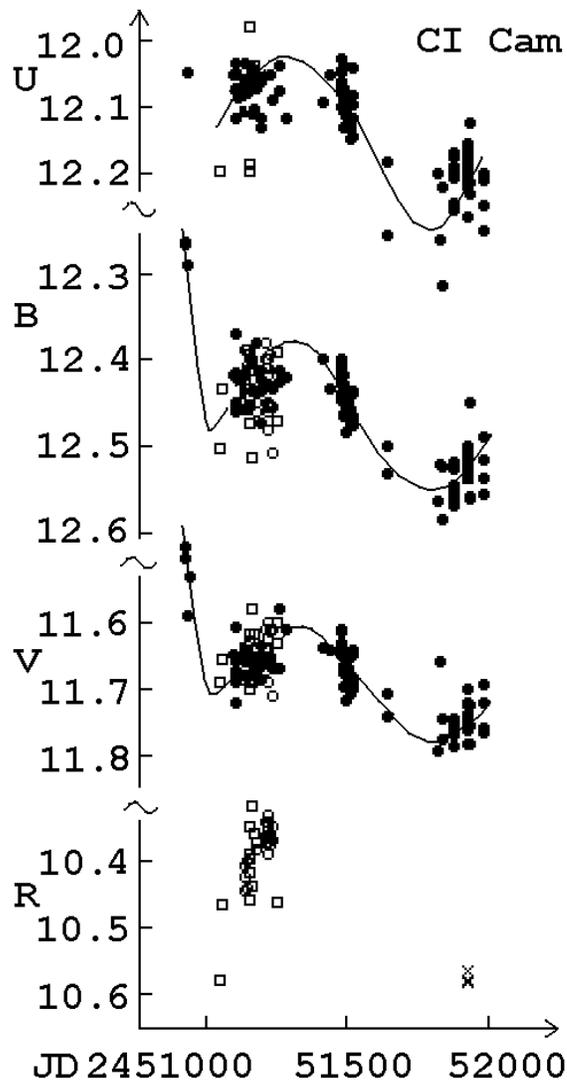,width=8cm}}\vspace{0.5cm}
\caption{Matter-dominated  evolution  of  $\varphi$  in  converging
models      with     $\varphi_0<0$             and
$\varphi^{\prime}_{0}=\varphi_{crit}$ (solid line),  $\varphi_0<0$ and
$\varphi^{\prime}_{0}>\varphi_{crit}$ (dotted line), $\varphi_0>0$ and
$\varphi^{\prime}_{0}=\varphi_{crit}$ (dashed line), and $\varphi_0>0$
and $\varphi^{\prime}_{0}>\varphi_{crit}$ (dotted-dashed  line), where
$\varphi_{crit}\equiv-(3/4+\beta)\varphi_0$. }
\label{fig1}
\end{figure}

This result can be understood by considering the scalar field $
\varphi $  as the position  of a particlelike dynamical variable which
moves under  a potential  $V(\varphi)$. Since  the right-hand side  of
Eq.   (\ref{Edecoupled}) represents a force   term,  this potential is
given by:
\begin{equation}\label{Gpotential}
V(\varphi)= (1-3\gamma)\int\alpha(\varphi)d\varphi
\end{equation}

If $ \alpha (\varphi)$ is  given by Eq. (\ref{Ecoupling}) with $\delta
= 1$, the matter-dominated form of $V(\varphi)$ becomes:
\begin{equation}\label{Potential1}
V(\varphi)=\frac{1}{2}\sigma_{\varphi}B_{2}\varphi^2
\end{equation}
where  we have  taken   the origin  of   potentials so that   $V=0$ at
$\varphi=0$.

Figure  2 shows this  potential.  We see  that $V(\varphi)$ has  a
stationary point at $\varphi=0$
which  does not  correspond to a   local minimum or maximum. Particles
with  $\varphi>0$  are  attracted
towards   $\varphi=0$ while those   with  $\varphi<0$ are rejected from
$\varphi=0$.

In the  case of ST theories  characterized by an  inefficient friction
($B_2>3/8$), any particle will 'slide' on this potential and will move
toward  $ \varphi\to   -\infty $, whatever     the initial values   of
$\varphi$ and  $\varphi'$ are. This  explains our previous result that
any theory with $B_2>3/8$ diverges from GR.

In the opposite case of theories  characterized by an efficient enough
friction ($B_2\le 3/8$), particles  with $\varphi>0$
have an overdamped  or  critically damped  motion. Therefore,   if the
initial  velocity $\varphi^{\prime}_{0}$ is   not more negative than a
critical value  $-(\beta_1+3/4)\varphi_0$,  the  energy  is completely
dissipated  by  the  friction   term,   and  the particle   stops   at
$\varphi=0$. On the contrary,  if $\varphi<0$, the tendency to diverge
from  GR ($\varphi=0$) can   only be overcome if  the  particle has an
initial velocity high enough to reach a point placed in the attraction
($\varphi>0$) region. This explains  our  previous result that, in  ST
theories  with   $B_2\le3/8$,  the convergence  towards  GR  impose on
$\varphi^{\prime}_{0}$        the      constraints       given      by
Eqs. (\ref{Ccondition}).

\begin{figure}
\centerline{\epsfig{figure=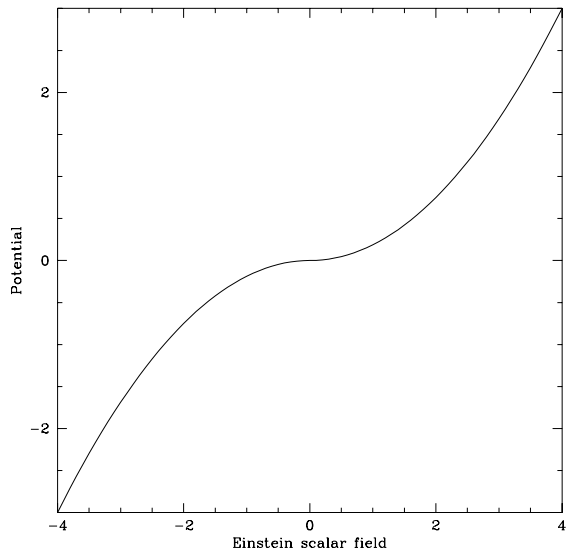,width=8cm}}\vspace{0.5cm}
\caption{Potential $ V(\varphi )$ for scalar-tensor theories with
$\delta =1$. Here, the  particular case $B_2=1/9$ is shown}
\label{fig2}
\end{figure}

\subsection{Case of a variable elastic coefficient ($\delta\ne 1$)}

When $\delta\ne 1$, the matter-dominated potential is given by
\begin{equation}\label{potential}
V(\varphi )=\frac{2-\delta}{2}\sigma _{\varphi}B_{2}|\varphi|^{2/(2-\delta )}
\end{equation}

Since the form of  this potential is  qualitatively similar to that of
Eq. (\ref{Potential1}), we can expect  for these models the same types
of behaviours as those found  in the $\delta=1$ case. Nevertheless, as
we  have seen  in  the previous  subsection, these behaviours strongly
depend  on the  relative importance of  the friction  term and  on the
initial value of $\varphi$ and  $\varphi'$. We can obtain some insight
on  the conditions  for convergence toward  GR  by considering the two
following limiting cases.

When   $1<\delta<2$,     the    elastic     coefficient   defined   by
Eq.       (\ref{elastic})           tends      to    zero           as
$\varphi\rightarrow0$. Consequently, in  the limit close enough to GR,
we  can neglect   the   term  $\sigma_\varphi  \kappa(\varphi)\varphi$    of
Eq.(\ref{master2}), so   that  the  scalar field  evolution   equation
becomes {\it friction dominated}:
\begin{equation}
\varphi''+\frac{3}{2}\varphi'=0
\label{case2}
\end{equation}

The solution of this equation is:
\begin{equation}
\varphi=C_1+C_2 e^{-\frac{3}{2}p}=\varphi_0+\frac{2}{3}\varphi^{\prime}_{0}
(1-e^{-\frac{3}{2}p})
\end{equation}
which   implies that  $\varphi$   vanishes at a   finite  time $p_1>0$
provided that:
\numparts
\begin{eqnarray}
\varphi^{\prime}_{0}>-\frac{3}{2}\varphi_0\quad (\varphi_0<0) \\
\varphi^{\prime}_{0}<-\frac{3}{2}\varphi_0\quad (\varphi_0>0)
\end{eqnarray}
\endnumparts

These expressions correspond to   the limit  $B_2\to 0$  (very  strong
friction)   of  Eq.  (\ref{conditions})     for  the $   \delta    =1$
case. Therefore, according to Eqs. (\ref{Ccondition}), theories with $
\delta >1$ will converge toward GR provided that

\begin{equation}\label{Ccondition2}
\varphi^{\prime}_{0}\ge-\frac{3}{2}\varphi_0
\end{equation}

On the contrary, when $1/2<\delta<1$,  the elastic coefficient becomes
infinity  as  $\varphi\rightarrow0$   (see    Eq. \ref{elastic}).  The
friction term of Eq.  (\ref{Edecoupled}) is then very inefficient and,
therefore, ST theories behave as the limit  $B_2\to \infty$ (very weak
friction) of the  $ \delta =1$ case.  That is,  the $\varphi$ particle
will 'slide' on the potential and will move towards $\pm\infty$. These
models  will then  diverge  from GR  whatever the  initial   values of
$\varphi$ and $\varphi'$ are.

\subsection{Influence of the mass term}

In the preceding discussion, we have obtained necessary and sufficient
conditions for convergence toward GR in  the limit of small $\varphi'$
velocities. We will now show that, in the  most general situation with
a  non-negligible $\varphi'$, there also  exists  a necessary (but not
sufficient) condition for convergence toward GR.

Let us consider Eq. (\ref{master}) with  the coupling function defined
in Eq. (\ref{Ecoupling}):
\begin{equation}
\frac{2}{3-\varphi'^2}\varphi''+\varphi'=
-\sigma_\varphi\kappa(\varphi)\varphi\label{master3}
\end{equation}

By integrating this equation, one obtains
\begin{eqnarray}
 \ln\left( \frac{\sqrt{3}+\varphi'}{\sqrt{3}+\varphi^{\prime}_{0}}\;
\frac{\sqrt{3}-\varphi^{\prime}_{0}}{\sqrt{3}-\varphi'}\right)
^{\frac{1}{\sqrt{3}}} &+&
(\varphi-\varphi_0)=\nonumber\\
&=& -B_2\int^{p}_{0}|\varphi|^{\frac{\delta}{2-\delta }} dp \label{Cmass}
\end{eqnarray}

Let  us now assume that theories  converge toward GR as $ p\rightarrow
+\infty      $.  In  this       case,   $\varphi\rightarrow   0$   and
$\varphi'\rightarrow   0$ as   $  p\rightarrow  +\infty   $,  so  that
Eq. (\ref{Cmass}) becomes
\begin{equation}
 \ln\left( \frac{\sqrt{3}+\varphi^{\prime}_{0}}{\sqrt{3}-
\varphi^{\prime}_{0}}\right)^{\frac{1}{\sqrt{3}}} +\varphi_0=
B_2\int^{\infty}_{0}|\varphi|^{\frac{\delta}{2-\delta }} dp >0
\end{equation}

Therefore, the  assumption  of   convergence  toward GR lead   to  the
necessary condition:
\begin{equation}
\varphi^{\prime}_{0}>\sqrt{3}\frac{1-e^{\sqrt{3}\varphi_0}}
{1+e^{\sqrt{3}\varphi_0}}
\end{equation}
which leads to $\varphi^{\prime}_{0}>-3\varphi_0/2$ when $\varphi_0\ll
1$ while, in   the limits $\varphi_0\rightarrow  \pm  \infty$, implies
$\varphi^{\prime }_{0}>\mp 1$

\section{Implications on Primordial Nucleosynthesis}\label{nucleosynthesis}

The  primordial  nucleosynthesis process  (PNP) starts,  in  the early
universe, soon  after the cosmic temperature  becomes lower  than that
needed   to maintain  the proton-to-neutron  ratio  in its equilibrium
value (freezing-out temperature). The nuclear reactions that then take
place lead  first   to the Deuterium    and Helium-3 formation.  These
elements are then burnt to produce  Helium-4 and very small amounts of
heavier elements. This process is sensitive  to the universe expansion
rate during nucleosynthesis, which can be parametrized by the speed-up
factor $\xi\equiv    H/H_{GR}$,  where $H$    is  the  physical Hubble
parameter and   $H_{GR}$  is   that  predicted  by   GR  at  the  same
temperature. Since scalar-tensor theories predict a universe expansion
rate which differs  from  that obtained  in  the framework of GR,  the
light-element production can be used to place constraints on the $\xi$
value at the beginning of the PNP process, $\xi_{PNP}$.

Most of works agree on  the PNP constraint on $\xi_{PNP}$  (subscripts
PNP  denote  values  at  the  beginning  of nucleosynthesis).  When  a
conservative  choice for  the   observed  primordial   abundances   is
considered, $\xi_{PNP}$ is  constrained to be $0.8 \lesssim  \xi_{PNP}
\lesssim 1.2$ \cite{DP99,Barrow78}. However, when a more severe choice
for  the observed abundances is taken,  the PNP  limits on $\xi_{PNP}$
are $0.95 \lesssim \xi_{PNP} \lesssim 1.03$ \cite{SA96b,Casas92}.  The
problem arises when this  bound  is used  to place constraints  on the
present value of the coupling function, $\omega_0$.

Serna \& Alimi \cite{SA96b}  have considered a family of scalar-tensor
theories where  the matter-dominated potential $V(\varphi)$ is similar
to that of Fig. 2 and, in  addition, the coupling function $3+2\omega$
is  {\it   a monotonic  function  of  time} (i.e.,
$3+2\omega$ diverges  to infinity  only in  the   limit of very  large
times).   Except   for    some  particular   cases   \cite{AS97},  the
nucleosynthesis bounds obtained  on these theories are very stringent:
$\omega_0\gtrsim  10^{20}$.  Other authors \cite{Santiago97,DP99} have
instead considered a different  family of theories where  $V(\varphi)$
has a constant  sign  and $3+2\omega$ is  not necessarily  a monotonic
function  of time. The  PNP bounds  on  these last  theories are about
thirteenth orders  of magnitude  lower ($\omega_0\gtrsim 10^{7}$) than
those obtained in the previous case.

It has been suggested in the  literature that the above discrepancy in
the PNP bounds on $\omega_0$ could be  perhaps due to the existence in
Ref.   \cite{SA96b}  of numerical  instabilities  (runaway  solutions)
arising    when  the  field   equations  are   solved  from a backward
time-integration. We  will now show that the  integration sense has no
effects on the  PNP bounds  (see the Appendix   A  for a test on   the
numerical stability against the time-integration sense). The important
point     to  explain  the  above    discrepancies  is  the particular
scalar-tensor theory used to describe the universe evolution.

We  will consider  the same  family  of  scalar-tensor theories as  in
Ref.  \cite{SA96b}, and  we will  restrict  our discussion to singular
models  with $3+2\omega >0$ and  a monotonic evolution of the speed-up
factor.

Let  us   first  consider  the   case  $\Phi>1$  ($\varphi<0$),  where
$\sigma_\varphi<0$ and the scalar field enters the matter-dominated era with
the overdamped repulsion  motion of Eq. (\ref{solution4}).   According
to  Eq. (\ref{Ccondition}),  the   convergence toward GR   requires an
initial  scalar field velocity,   $ \varphi^{\prime}_{M} $ (subscripts
$M$ denote values at the  beginning of the matter-dominated era), with
a positive value:
\begin{equation}\label{MDcondit}
\varphi^{\prime}_{M}\geq -\left[\frac{3}{4}+\beta_2  \right] \varphi_M
\end{equation}

Therefore, since  $\varphi_M<0$, the choice $ \varphi^{\prime}_{M}=0 $
is {\it  forbidden}  for this  family   of scalar-tensor  theories. In
addition,   the requirement of   a  monotonic time  evolution for  the
coupling function eliminates the possibility of $\varphi^{\prime}_{M}>
-(3/4+\beta_2)  \varphi_M$,  where convergence   toward GR is obtained
after a rather complicate behaviour of $3+2\omega$ in which it becomes
infinity (when $\varphi=0$  for the  first  time), then  it  decreases
until reaching a local minimum value (maximum value of $\varphi$), and
finally it increases monotonically to infinity (see Fig. 2).

Taking the only possible choice for these theories
\begin{equation}\label{goodchoice}
\varphi^{\prime}_{M}= -\left[\frac{3}{4}+\beta_2  \right] \varphi_M,
\end{equation}
Eq. (\ref{solution4}) reduces to:
\begin{equation}\label{MDvarphi}
\varphi= \varphi_M e^{-(\frac{3}{4}+\beta_2)p }
\end{equation}
where,  from Eqs. (\ref{constant4a})-(\ref{constant4b}), $  \beta_2 $  has a constant value
larger than  3/4  (very  overdamped  motion, $B_2\rightarrow0$)    and
smaller   than    $     3\sqrt{2}/4$   (critically     damped  motion,
$B_2\rightarrow3/8$).

Using Eqs. (\ref{Ecoupling}) and  (\ref{MDvarphi}), the present  value
of the coupling function is (with $ \delta =1$):
\begin{equation}\label{alpha2}
\alpha^{2}_{0}= B^{2}_{2}\varphi^{2}_{M} e^{-2(\frac{3}{4}+\beta_2)p_{0} }
\end{equation}
where $p_0\approx 10$ (see Ref. \cite{DN93}) is the amount of $p$ time
elapsed since the beginning  of  the matter-dominated era.  Taking the
less stringent value $B_2\approx 1/9$, Eq. (\ref{alpha2}) implies
\begin{equation}
\alpha^{2}_{0}\approx   10^{-16} \varphi^{2}_{M}
\end{equation}
while using the largest value $B_2=3/8$
\begin{equation}
\alpha^{2}_{0}\approx   10^{-17} \varphi^{2}_{M}
\end{equation}
In order to estimate $\alpha^{2}_{0}$,  we must specify a  $\varphi_M$
value compatible with  the  observed abundance  of light elements.  To
that end      we  note that,  according   to    Eqs. (\ref{pdef})  and
(\ref{Hast}),  the physical  Hubble parameter  $H$ is  related to that
measured in Einstein units $H_{\ast}$ by
\begin{equation}
H^2=\frac{H^{2}_{\ast}}{A^2}(1+\alpha \varphi')^2
\end{equation}
Therefore, using Eq. (\ref{Hast2}), we obtain
\begin{equation}
H^2=\frac{8\pi G\rho A^2}{3-\varphi'^2}(1+\alpha \varphi')^2
\end{equation}
and the speed-up factor  $\xi\equiv H/H_{GR}$  (where $H^{2}_{GR}=8\pi
G\rho /3$) is given by:
\begin{equation}
\xi ^2=\frac{3 A^2}{3-\varphi'^2}(1+\alpha \varphi')^2
\end{equation}
which,  in  the limit  of  very small $\varphi'$  values, implies $\xi
\simeq A$

In  models with  a monotonic  evolution of   the speed-up factor,  the
compatibility with  the observed abundance   of light elements imposes
$0.95\lesssim    \xi  \lesssim    1.03$   just   before the   big-bang
nucleosynthesis process.  Since $\xi \simeq  A$  and $\psi=-lnA$, this
constraint           implies         (see       Eq.      \ref{varphi})
:\begin{equation}\label{phippn}
\varphi_{PPN}\approx \left\{
\begin{array}{ll}
-0.72 & (\mbox{if } B_2=1/9) \\
-0.39 & (\mbox{if } B_2=3/8)
\end{array}\right.
\end{equation}

Taking into account  the condition  (\ref{goodchoice}) for convergence
toward GR (i.e., $\varphi^{\prime}_{M}\approx -\varphi_{M}$ at the end
of the  radiation  era),  Eq.   (\ref{RDvelocity}) implies   $k\approx
-\varphi_{M}e^{p_R}/\sqrt{3} $ and Eq. (\ref{RDvarphi}) then yields
\begin{equation}\label{PPN}
\varphi_{PPN}\approx \varphi_{M}-\sqrt{3}\ln
\left[\frac{\sqrt{\varphi^{2}_{M}e^{2p_R}+3}-\varphi_M e^{p_R}}
{\sqrt{\varphi^{2}_{M}+3}-\varphi_M} \right]
\end{equation}
where   $p_R$ denotes the  $p$-time  elapsed  from the relevant epochs
prior to the big  bang nucleosynthesis up to  the end of the radiation
era. Using $\varphi_{PPN}$ values of Eq. (\ref{phippn}) and the rather
conservative value $p_R=3$ (i.e., the  cosmic temperature decreases in
about three orders of magnitude) , Eq. (\ref{PPN}) implies:
\begin{equation}
\varphi_M\lesssim
 \left\{
\begin{array}{ll}
-3.7\cdot 10^{-2} & (\mbox{if } B_2=1/9) \\
-2.0\cdot 10^{-2}& (\mbox{if } B_2=3/8)
\end{array}\right.
\end{equation}
Consequently,
\begin{equation}\label{bound}
\alpha^{2}_{0}\lesssim    \left\{
\begin{array}{ll}
10^{-19} & (\mbox{if } B_2=1/9) \\
10^{-21}& (\mbox{if } B_2=3/8)
\end{array}\right.
\end{equation}
in good  accord with the  numerical  computations of  Serna and  Alimi
\cite{SA96b}.

The case $\Phi<1$ ($\varphi>0$)  requires a different analysis.  Since
the  models $ 3+2\omega >0$ and   $\Phi<1$ studied in \cite{SA96b} are
all nonsingular (class-3), the cosmic temperature has a maximum value,
$T_{max}$,  which    corresponds   to  the   minimum   of    the scale
factor.  Consequently, in    this class  of  models,   the  bounds  on
$\alpha_0$ are mainly imposed by the  condition that $T_{max}$ must be
high enough  to    allow  for the   existence   itself of   primordial
nucleosynthesis. The  numerical computations  of \cite{SA96b} lead  to
limits  on  $\alpha^{2}_{0}$        close   to  those    given      by
Eq. (\ref{bound}).  This  nonsingular  behaviour   is found for    any
$\varphi^{\prime}_{M}$  value, except   for  the (fine  tuned)  choice
$\varphi^{\prime}_{M}=0$ where the scalar field evolution is frozen to
a constant value during the radiation epoch.

It is  worth commenting  the fact  that the strong  limit expressed in
Eq.  (\ref{bound}) refers  to  scalar-tensor models  with a  monotonic
evolution of $\xi(T)$. As  we remarked in previous works \cite{SA96b},
these limits lead  to cosmological models  which do not  significantly
differ from the  standard GR one except  perhaps for very early epochs
before nucleosynthesis. On the contrary,  when the speed-up factor has
not a monotonic  behaviour during  the  big-bang nucleosynthesis,  the
compatibility    with  the observed abundance    of  light elements is
possible even for very  large $\xi_{PPN}$ values \cite{SA96b,AS97}. In
the framework   of  this last family   of  models, the nucleosynthesis
constraints   on $\varphi_{M}$   and  $\alpha^{2}_{0}$  are  not  very
stringent and,  in addition, the allowed  range for the baryon density
is much wider than in the standard GR cosmologies.

\section{Conclusions}\label{conclusions}

In  this  paper we  have  analyzed  the  convergence of  scalar-tensor
theories (ST)  toward GR and  its consequences on  the nucleosynthesis
bounds on the present value of the coupling function.

To that end, we have  deduced an autonomous evolution equation for the
Jordan scalar field.   By writing this equation in  Einstein units, we
have analyzed the evolution of the
scalar  field  both  in  the  radiation-dominated  epoch  and  in  the
matter-dominated  epoch.   We  have  considered  a  coupling  function
defined  by Eq.   (\ref{Ecoupling}), which  reproduces all  the models
proposed by Barrow \& Parsons \cite{BP97} in the limit close to GR. We
have then shown that, in general, the evolution of the scalar field is
governed  by two opposite  mechanisms: an  attraction and  a repulsion
mechanism. The attraction mechanism dominates the recent epochs of the
universe  evolution  if,  and  only  if,  the  scalar  field  and  its
derivative satisfy  certain boundary  conditions which depend  on each
particular scalar-tensor theory.

Our results have been then applied to obtain an analytical estimate of
the  Big-Bang nucleosynthesis  (BBN)  bounds on  $\omega_0$.  We  have
found  that the  particular ST  theory used  to describe  the universe
evolution   has  a   crucial   importance  on   the   BBN  limits   on
$\omega_0$. Therefore, it  is not possible to establish  a general and
unique  limit  for all  ST  models.  In  the  particular  case of  the
theories analyzed in this paper, where $\alpha \propto |\varphi|$, our
analytical estimates  are in close agreement  with the nucleosynthesis
bounds  numerically obtained in  \cite{SA96b} ($\alpha^{2}_{0}\lesssim
10^{-20}$).   Theories different  from  those analyzed  in this  paper
could imply very different BBN bounds.  For instance, in the case of a
ST theory defined by $\alpha(\varphi) \propto \varphi$, where only the
attraction    mechanism    is   present,    the    BBN   bounds    are
($\alpha^{2}_{0}\lesssim  10^{-7}$)  \cite{SKW99,DP99}.   In the  same
way, in  scalar-tensor theories with a non-monotonic  evolution of the
speed-up factor,  the BNB  limits are drastically  relaxed to  a value
comparable   to   that   obtained   from  solar   system   experiments
($\alpha^{2}_{0} \lesssim 0.02$).  In addition, in this last case, the
allowed  range for  the  baryon density  is  much wider  than in  the
standard GR cosmologies \cite{SA96b,AS97}.

\ack{This  work  has  been  partially  supported  by  the
Generalitat Valenciana (project number GV00-139-1), Spain.}

\section*{Appendix A: Numerical stability of the time-integration
sense}
\label{section-stability}

We will here comment a series of computations which were performed in
\cite{SA96a,SA96b}  to  test  the  numerical  stability   against  the
time-integration sense.  In  these computations, the  ST  cosmological
equations were solved in the two senses (forward and backward in time)
by using a sixth-order Runge-Kutta  method which takes the temperature
as variable. A standard particle content (baryons, electron-positrons,
photons and neutrinos) of the  universe is assumed, so that transition
phases are solved without approximations.

\begin{figure}
\centerline{\epsfig{figure=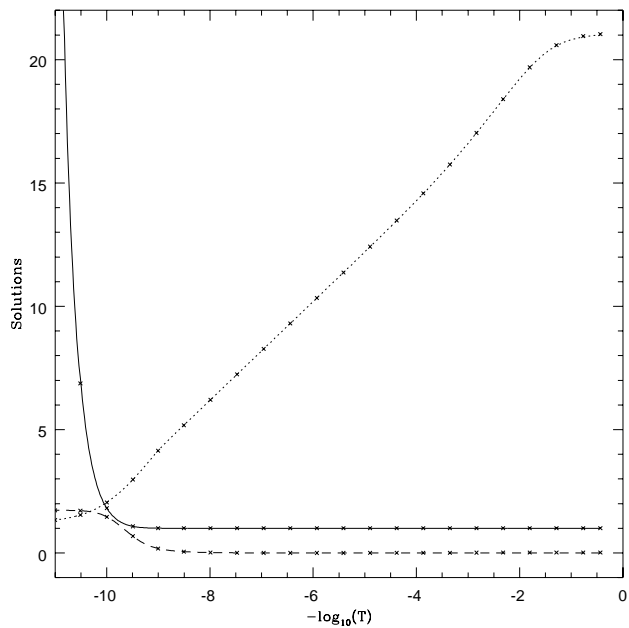,width=9cm}}
\caption{Numerical solutions for $\xi$ (solid line),
$\log_{10}[1+\alpha^{-2}]$ (dotted line)  and $\varphi'$ (dashed line)
in an  open ($k=-1$)  scalar-tensor   theory defined  by  $\alpha^2  =
(\lambda^2/3)  (1-\Phi)$ with  $\lambda^2=0.9$.   The results obtained
from  a backward  and a  forward time-integration  are represented  by
lines and crosses, respectively. }
\label{fig3}
\end{figure}

Figure \ref{fig3}  shows a  typical  example of the numerical  results
found for the    family   of scalar-tensor  theories    considered  in
\cite{SA96a,SA96b}.  The lines in  this figure represent the solutions
obtained for $\xi$   (solid line), $\log_{10}(1+\alpha^{-2})$  (dotted
line)  and $\varphi'$ (dashed  line) from a backward time-integration.
The  final results  of   this integration   were   taken as   boundary
conditions to a reverse (forward  in  time) integration.  The  results
obtained in  this last  case are  represented by crosses  in the  same
figure.

As we can see from Fig. 3, both integration senses give almost exactly
the same  results.   This fact   can  be very well  understood  if one
analyses    the    propagation   of   numerical   errors   during  the
integration. To that end, we consider the simple  case of a critically
damped  motion ($B_2 =  \frac{3}{8}$),   where the time evolution   of
$\varphi$ is given by Eq. (\ref{solution2}):
\begin{equation}
\varphi(p) = \varphi_0
\left[1+\left( \frac{\varphi'_0}{\varphi_0}+\frac{3}{4}\right)p\right]
e^{-\frac{3}{4}p}
\end{equation}

Let now suppose that we integrate  this equation backward in time ($p$
varying from 0 today up  to large negative values   in the past)  with
numerical errors  $\Delta   \varphi_0  \ll  \varphi_0$   and   $\Delta
\varphi'_0 \ll \varphi'_0$   on the initial  values.  Obviously, since
the   functions $\varphi$ and  $\varphi'$  increase in the past, their
absolute errors will also  increase. However the relevant  quantity to
estimate the accuracy of the integration is not the absolute error but
the relative error which, at any $p$-time, is given by
\begin{equation}
\left(\frac{\Delta \varphi}{\varphi}\right)(p) =
\frac{\Delta \varphi_0 + (\frac{3}{4}\Delta \varphi_0 +
\Delta \varphi'_0)p}
{\varphi_0 + (\frac{3}{4}\varphi_0 + \varphi'_0)p}
\end{equation}
\begin{equation}
\left(\frac{\Delta \varphi'}{\varphi'}\right)(p) =
\frac{-\Delta \varphi'_0 +
\frac{3}{4}(\frac{3}{4}\Delta \varphi_0 + \Delta \varphi'_0)p}
{-\varphi'_0 + \frac{3}{4}(\frac{3}{4}\varphi_0 + \varphi'_0)p}
\end{equation}
In the limit where $p$ goes to $-\infty$, the above expressions become
\begin{equation}
\left(\frac{\Delta \varphi}{\varphi}\right)_{-\infty} =
\left(\frac{\Delta \varphi'}{\varphi'}\right)_{-\infty} =
\frac{\frac{3}{4}\Delta \varphi_0 + \Delta \varphi'_0}
{\frac{3}{4}\varphi_0 + \varphi'_0}
\end{equation}
Since $\Delta \varphi_0 \ll \varphi_0$ and $\Delta \varphi'_0 \ll
\varphi'_0$, we then deduce that the  relative errors remain very small in
the  past.  This explains why the  computation presented in Fig. 3 are
stable in both integration senses.

\end{document}